\documentclass{desyproc}

\usepackage{units}
\newcommand{\pt}{p_\mathrm{T}}

\begin{document}
\title{Direct Photons at RHIC}

\author{{\slshape Klaus Reygers$^1$ for the PHENIX Collaboration}\\[1ex]
$^1$Physikalisches Institut, Universit{\"a}t Heidelberg, Philosophenweg 12, 69120 Heidelberg, \\Germany}

\contribID{75}

\confID{1407}  
\desyproc{DESY-PROC-2009-03}
\acronym{PHOTON09} 
\doi  

\maketitle

\begin{abstract}
  A brief overview of direct-photon measurements in p+p and Au+Au
  collisions at $\sqrt{s_{NN}} = \unit[200]{GeV}$ with the PHENIX
  experiment at the Relativistic Heavy Ion Collider (RHIC) is
  given. Direct-photon yields for $\pt \gtrsim \unit[4]{GeV}/c$ and
  photon-hadron azimuthal correlations were determined with the aid of
  an electromagnetic calorimeter.  By detecting $e^+e^-$ pairs from
  the internal conversion of virtual photons direct-photon yields were
  measured between $1 \lesssim \pt \lesssim \unit[4]{GeV}/c$. In Au+Au
  collisions thermal photons from a quark-gluon plasma (QGP) are
  expected to contribute significantly to the total direct-photon
  yield in this range.
\end{abstract}

\section{Introduction}
\label{Sec:intro}
In heavy-ion physics direct photons are typically defined as the
difference between all measured photons and background photons from
hadronic decays \cite{Stankus:2005eq}. Thus, isolated prompt photons
with small hadronic activity around them accompanied by a jet on the
away-side as well as photons produced in the fragmentation of jets
(fragmentation photons) contribute to the direct-photon signal. The
primary reason for the interest in direct photons is their large mean
free path with respect to the dimensions of the created
fireball. Thus, once produced photons leave the fireball unscathed and
carry away information about the early stage of the collisions.

The measured direct-photon signal is an integral over the entire
evolution of the fireball where different processes are dominant at
different times. This is often regarded as a virtue, however, it also
means that disentangling the different sources typically relies on
comparisons with model calculations. The production of direct photons
in ultra-relativistic A+A collision can be divided into the following
stages \cite{Gale:2009gc}. At first, direct photons are produced in
initial hard parton scatterings analogous to the production mechanism
in p+p collisions. The yield of these photons can be calculated in
perturbative QCD and they are the dominant direct-photon source at
high $\pt$ ($\pt \gtrsim \unit[6]{GeV}/c$ for Au+Au at $\sqrt{s_{NN}}
= \unit[200]{GeV}$). After a time on the order of $\tau_0 \approx
\unit[1]{fm}/c$ \cite{Kolb:2003dz} it is expected that a medium of
deconfined quarks and gluons (the quark-gluon plasma) forms for which
a local temperature is a meaningful concept. In such a thermalized
medium thermal direct photons will be produced whose momentum
distribution reflects the temperature of the system.  At a temperature
of $T_c \approx \unit[140 - 200]{MeV}$ \cite{Aoki:2009sc,Cheng:2006qk}
a transition to a hot hadron gas takes place and thermal direct
photons are also produced in this phase.

It was discovered at RHIC that quark and gluon jets in central A+A
collisions are affected by the created medium. Jets apparently lose
energy which results, {\it e.g.}, in a reduced yield of pions at high
$\pt$ \cite{Adcox:2004mh}. This is referred to as jet quenching. The
jet-medium interaction gives rise to further sources of direct
photons. First, a direct photon can be produced in so-called
jet-photon conversions, {\it e.g.}, in gluon Compton scattering
$q_\mathrm{jet} + g_\mathrm{QGP} \rightarrow q + \gamma$
\cite{Fries:2002kt}. In these processes the photon typically carries a
large fraction of the initial jet energy. Second, the presence of the
medium induces the emission of bremsstrahlung photons
\cite{Zakharov:2004bi}. This is analogous to the induced gluon
emission which is believed to be the dominant mechanism for the jet
energy loss.

\section{Direct Photons at High \boldmath $\pt$ \unboldmath}

\begin{figure}[t]
\centerline{\includegraphics[width=\textwidth]{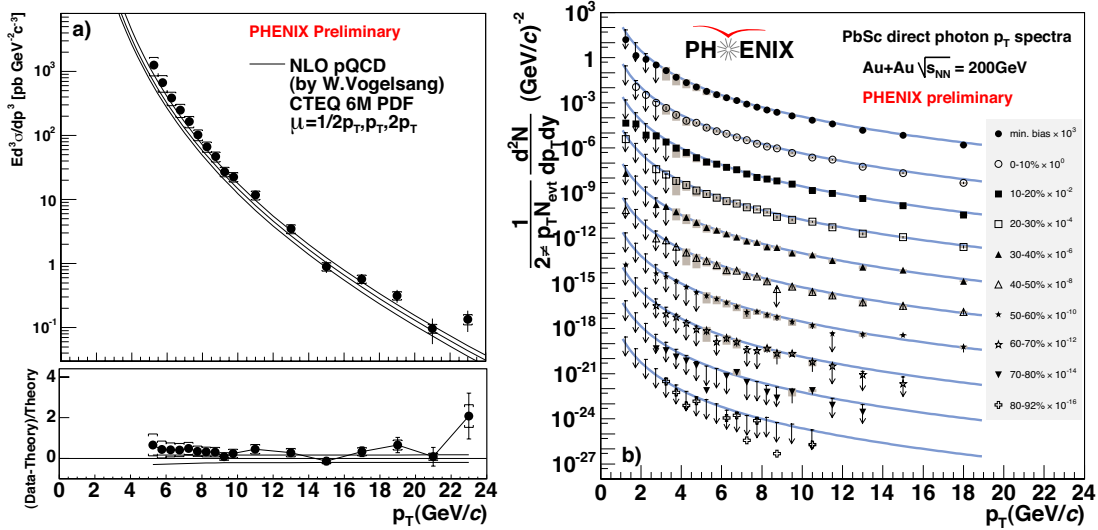}}
\caption{a) Direct-photon invariant cross section in p+p collisions at
  $\sqrt{s} = \unit[200]{GeV}$ from the 2005 run. The data agree with
  a next-to-leading-order (NLO) QCD calculation.  Final data from 2003
  run were published in \cite{Adler:2006yt}. b) Direct-photons yields
  in Au+Au collisions at $\sqrt{s_{NN}} = \unit[200]{GeV}$ from the
  2004 run for various centralities.  The data are compared a p+p NLO
  QCD calculation scaled by $\langle T_\mathrm{AB} \rangle = \langle
  N_\mathrm{coll} \rangle / \sigma_\mathrm{inel}^\mathrm{NN}$.  Final
  results from the 2002 run were published in \cite{Adler:2005ig}.
}\label{Fig:photon_spectra_high_pt}
\end{figure}

In the PHENIX experiment direct photons at midrapidity ($|y| < 0.35$)
above $\pt \approx \unit[4]{GeV}/c$ are measured with an
electromagnetic calorimeter (EMCal) \cite{Aphecetche:2003zr}. This
detector subtends $\Delta \phi \approx \pi$ in azimuth and consists of
highly segmented lead-scintillator sampling (PbSc, 6 sectors) and
lead-glass Cherenkov calorimeters (PbGl, 2 sectors).  The two detector
technologies have different systematics and provide the possibility of
internal cross-checks. In the Au+Au analysis the ratio
$(\gamma_\mathrm{inclusive}/\pi^0)_\mathrm{meas}$ of the inclusive
photon spectrum, {\it i.e.}, the spectrum of photon from all sources
including decay photons, and the $\pi^0$ spectrum is calculated.  A
direct-photon excess can then be found be dividing this ratio by
$(\gamma_\mathrm{decay}/\pi^0)_\mathrm{calc}$, {\it i.e.}, by the
calculated number of hadronic decay photons per $\pi^0$.  The dominant
contribution to these background photons comes from $\pi^0\rightarrow
\gamma\gamma$ and $\eta\rightarrow \gamma\gamma$.  Extracting the
direct-photon excess from the ratio
$(\gamma_\mathrm{inclusive}/\pi^0)_\mathrm{meas}$ has the advantage
that uncertainties of the energy scale of the calorimeter partially
cancel. The direct-photon yield is then calculated as
\begin{equation}
\gamma_\mathrm{direct} = \gamma_\mathrm{inclusive} - \gamma_\mathrm{decay}
= \left(1 - R^{-1} \right) \times \gamma_\mathrm{inclusive}
\quad \mathrm{with} \quad
R = \frac{\left( \gamma_\mathrm{inclusive}/\pi^0 \right)_\mathrm{meas}}
{\left( \gamma_\mathrm{decay}/\pi^0 \right)_\mathrm{calc}} \;.
\end{equation}
In p+p collisions also a slightly different statistical subtraction
method is employed \cite{Adler:2006yt}.

Figure \ref{Fig:photon_spectra_high_pt}a shows that the measured
invariant direct-photon cross section in p+p collisions at $\sqrt{s} =
\unit[200]{GeV}$ agrees with a next-to-leading-order (NLO) QCD
calculation.  In the absence of nuclear effects yields of hard
scattering processes are expected to scale as $\langle T_\mathrm{AB}
\rangle \times E\,\mathrm{d}^3\sigma/\mathrm{d}^3p|_{p+p}$.  The
nuclear overlap function $\langle T_\mathrm{AB} \rangle$ reflects the
nuclear geometry and is related to the number of inelastic
nucleon-nucleon collisions according to $\langle T_\mathrm{AB} \rangle
= \langle N_\mathrm{coll} \rangle / \sigma_\mathrm{inel}^\mathrm{NN}$
where $\sigma_\mathrm{inel}^\mathrm{NN}$ is the inelastic
nucleon-nucleon cross section \cite{Miller:2007ri}.  The scaled p+p
NLO QCD cross section agrees well with the measured direct-photon
yields as can be seen in Figure~\ref{Fig:photon_spectra_high_pt}b.

\begin{figure}[t]
\centerline{\includegraphics[width=\textwidth]{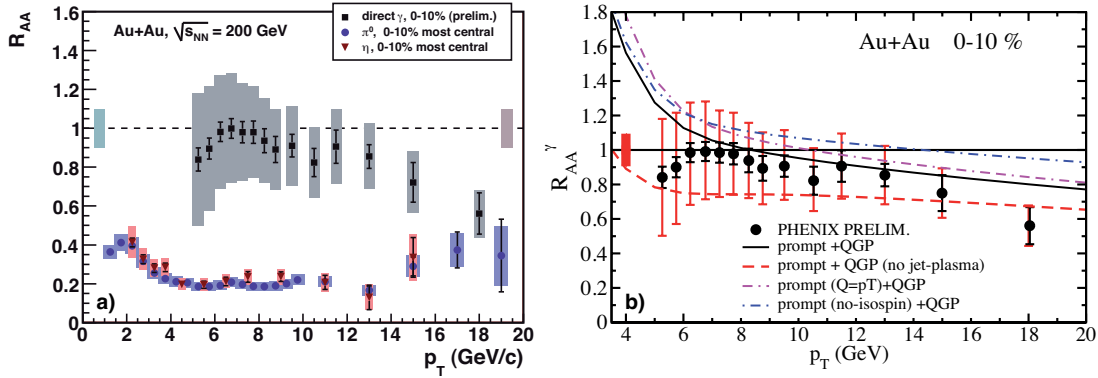}}
\caption{a) $R_\mathrm{AA}$ in central Au+Au collisions at
  $\sqrt{s_\mathrm{NN}} = \unit[200]{GeV}$ for direct photons,
  $\pi^0$'s and $\eta$'s. b) Direct-photon data from a) compared to a
  calculation which as a net result of different nuclear effects
  discussed in the main text predicts $R_\mathrm{AA} \approx 0.8$ at
  $\pt \approx \unit[20]{GeV}/c$.  }\label{Fig:Raa}
\end{figure}
The nuclear modification factor 
\begin{equation}
R_\mathrm{AB}(\pt) = \frac{\mathrm{d}N/\mathrm{d}\pt|_{A+B}}
{\langle T_\mathrm{AB} \rangle \times \mathrm{d}\sigma/\mathrm{d}\pt|_{p+p}}
\end{equation}
is used to quantify nuclear effects on the single particle yields. The
suppression of the $\pi^0$ and $\eta$ yields by a factor of $\sim 5$
in central Au+Au collisions at $\sqrt{s_{NN}} = \unit[200]{GeV}$
\cite{Adcox:2004mh} as shown in Figure~\ref{Fig:Raa}a is interpreted
as the result of energy loss of quark and gluon jets in a medium of
high color-charge density.  The yield of direct photons from initial
hard parton scatterings in A+A collisions is expected to scale with
$\langle T_\mathrm{AB} \rangle$ which is indeed observed in the region
$6 \lesssim \pt \lesssim \unit[12]{GeV}/c$. Thus, the direct-photon
results at high $\pt$ support the parton-energy loss interpretation.

At first sight the decrease of the direct photon $R_\mathrm{AA}$ below
unity for $\pt \gtrsim \unit[14]{GeV}/c$ spoils the simple picture of
the last paragraph. However, $\langle T_\mathrm{AB} \rangle$ scaling
of direct-photon yields is clearly an oversimplification. First, a
Au+Au collision can be regarded as a superposition of p+p, p+n, and
n+n collisions whereas only p+p collisions are used as reference in
the calculation of $R_\mathrm{AA}$. This so-called isospin effect
reduces $R_\mathrm{AA}$ at high $\pt$.  Moreover, the energy loss of
jets will lead to a reduced production of fragmentation photons. On
the other hand, anti-shadowing of the parton distribution in the Au
nucleus and bremsstrahlung photons from jet-plasma interactions will
increase $R_\mathrm{AA}$. In the calculation in Figure~\ref{Fig:Raa}
the combination of these effects results in an $R_\mathrm{AA} \approx
0.8$ at $\pt = \unit[20]{GeV}/c$. The experimental issue here is the
correction for the merging of the two showers from $\pi^0 \rightarrow
\gamma\gamma$.  A detailed study of this effect will be carried out
for the final publication.

\section{Photon-Triggered Away-side Correlations}
The pion yield at high $\pt$ in a given bin at $\pt^{\pi}$ results
from jets with a large spread in transverse momentum $\pt^\mathrm{jet}
\gtrsim \pt^{\pi}$. Thus, the measured pion $R_\mathrm{AA}$ contains
only indirect information about the energy loss of a jet with a given
energy. To better constrain the initial jet energy one can study jets
opposite ($\Delta \phi \approx \pi$) to a direct photon as for leading
order processes $\pt^\mathrm{jet} = \pt^\gamma$.  Full jet
reconstruction is difficult in heavy-ion reactions so that
photon-triggered away-side correlations are a useful tool to study jet
energy loss. One defines $z_\mathrm{T} =
\pt^\mathrm{hadron}/\pt^\gamma$ and the distribution $D(z_\mathrm{T})
= 1/N_\gamma^\mathrm{trig} \,
\mathrm{d}N^\mathrm{hadron}/\mathrm{d}z_\mathrm{T}$ approximates the
light quark fragmentation function \cite{Adare:2009vd}.

\begin{figure}[t]
\centerline{\includegraphics[width=\textwidth]{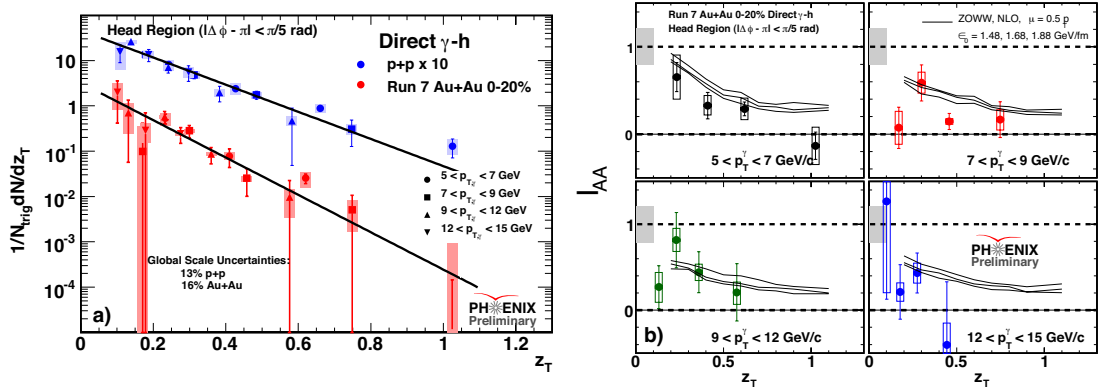}}
\caption{a) Charged-hadron yields opposite to a direct photon as a
  function of $z_\mathrm{T} = \pt^\mathrm{hadron}/\pt^\gamma$ in p+p
  (run 2005 + 2006) and Au+Au (run 2007) at
  $\sqrt{s_{NN}}=\unit[200]{GeV}$. b) The ratio $I_\mathrm{AA} =
  D_\mathrm{A+A}(z_\mathrm{T}) / D_\mathrm{p+p}(z_\mathrm{T})$ for
  four ranges the trigger photon $\pt$. Results from 2004 Au+Au run
  are published in \cite{Adare:2009vd}.}\label{Fig:gamma_hadron}
\end{figure}

The $z_\mathrm{T}$ distributions of charged hadrons associated with a
direct photon are shown in Figure~\ref{Fig:gamma_hadron}. If the
$z_\mathrm{T}$ distribution in p+p collisions is a good approximation
of the fragmentation function the distribution should scale in
$z_\mathrm{T}$, {\it i.e.}, it should only depend on $z_\mathrm{T}$
independent of the $\pt$ of the trigger photons. This is approximately
satisfied in p+p, but interestingly apparently also in Au+Au. The
distributions in p+p and Au+Au are fit with an exponential $\exp(-b
z_\mathrm{T})$. The difference between p+p ($b = 6.89 \pm 0.64$) and
Au+Au ($b = 9.49 \pm 1.37$) reflects the energy loss in the medium.

The ratios $I_\mathrm{AA} = D_\mathrm{A+A}(z_\mathrm{T}) /
D_\mathrm{p+p}(z_\mathrm{T})$ for different $\pt$ ranges of the
trigger photon are shown in Figure~\ref{Fig:gamma_hadron}.  They are
compared with a jet quenching calculation \cite{Zhang:2009rn} in which
an energy loss parameter was tuned to describe the single particle
$R_\mathrm{AA}$. Overall a good agreement with the data is observed.
However, the uncertainties of the data points are currently too large to
confirm the change of $I_\mathrm{AA}$ with the $\pt$ of the photon
trigger as predicted by the calculation.

\section{Direct Photons at Low \boldmath $\pt$ \unboldmath}
Systematic uncertainties related to the energy scale, the correction
of detector effects and the extraction of the $\pi^0$ yields prevent
the measurement of direct photons with the PHENIX EMCal below $\pt
\lesssim \unit[4]{GeV}/c$. This is the range in which the contribution
from thermal direct photons is expected to be largest.  A solution to
this quandary is the measurement of virtual photons with small mass
via their internal conversion in $e^+e^-$ pairs \cite{phenix:2008fqa}.
Electrons and positrons are identified within PHENIX with an Ring
Imaging Cherenkov detector and by matching the measured track momentum
with the energy signal in the EMCal. $e^+e^-$ pairs from external
conversions in the detector material are removed by a cut on the
orientation of the pair in the magnetic field. The combinatorial
background is subtracted using a mixed-event technique.  The remaining
correlated background is subtracted with the aid of constructing
like-sign pairs.

\begin{figure}[t]
\centerline{\includegraphics[width=\textwidth]{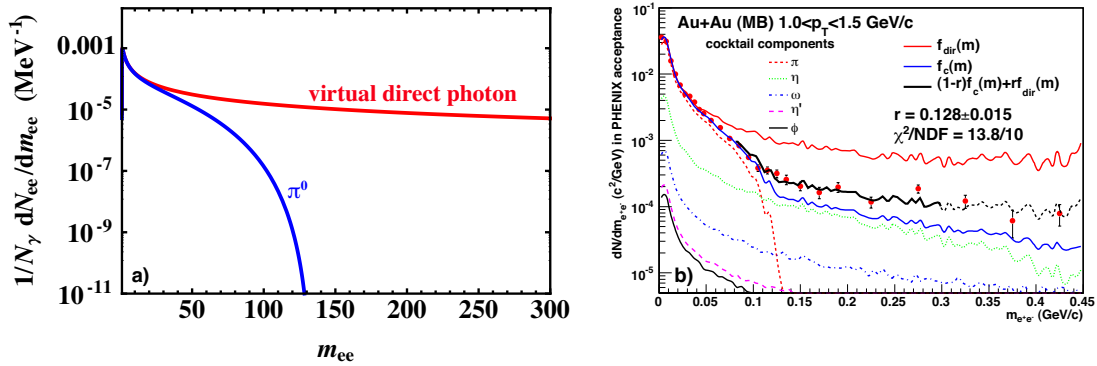}}
\caption{a) Mass distribution of $e^+e^-$ pairs from internal
  conversion as given by Equation~\ref{eq:Kroll-Wada}. b) Measured
  $e^+e^-$ mass distribution in Au+Au collisions at $\sqrt{s_{NN}} =
  \unit[200]{GeV}$. The data are well described by adding
  contributions from hadronic decays and virtual direct
  photons.}\label{Fig:mee}
\end{figure}

The internal conversion method exploits the fact that any source of
real photons also is a source of virtual photons and that the rate of
internal conversions and the mass distribution of the $e^+e^-$ pairs
is calculable within QED.  The number of $e^+e^-$ pairs per real
photon is given by \cite{Kroll:1955zu}
\begin{equation}
\frac{1}{N_\gamma}\frac{\mathrm{d}N_{ee}}{\mathrm{d}m_{ee}} = 
\frac{2\alpha}{3\pi}\frac{1}{m_{ee}} \sqrt{1-\frac{4m_e^2}{m_{ee}^2}}
\Bigl( 1+\frac{2m_e^2}{m_{ee}^2} \Bigr) S \;.
\label{eq:Kroll-Wada}
\end{equation}
For hadron decays, {\it e.g.}, the $\pi^0$ Dalitz decay, $S =
|F(m_{ee}^2)|^2 \left(1-m_{ee}^2/M_\mathrm{h}^2\right)^3$ where
$F(m_{ee}^2)$ is the form factor and $M_\mathrm{h}$ the hadron mass.
For a point-like process such as gluon Compton scattering ($q+g
\rightarrow q + \gamma^* \rightarrow q + e^+ + e^-$) $S \approx 1$ for
$\pt^{ee} \gg m_{ee}$. The two cases are shown in
Figure~\ref{Fig:mee}a.

At small masses $m_{ee} < \unit[30]{MeV}$ the mass distribution is to
very good approximation independent of the source and the fraction of
real direct photons can be expressed in terms of virtual photons, {\it
  i.e.}, $r \equiv \gamma_\mathrm{direct} / \gamma_\mathrm{inclusive}
= (\gamma^*_\mathrm{direct} / \gamma^*_\mathrm{inclusive})_{m_{ee} <
  \unit[30]{MeV}}$. At larger masses $e^+e^-$ pairs from hadronic
decays are suppressed by the $S$ factor and $m_{ee} < M_\mathrm{h}$
holds. Thus, the background from $\pi^0$ Dalitz decays could be
completely avoided by measuring virtual direct photons in the range
$m_{ee} > M_{\pi^0}$.  However, this comes at the expense of a loss in
statistics as for every real direct photon there are only $\sim 0.001$
virtual direct photons with $m_{ee} > M_{\pi^0}$.

For a given $\pt$ bin the direct-photon fraction $r$ is determined as
follows. First the mass distribution of $e^+e^-$ pairs from hadronic
decays $f_\mathrm{cocktail}$ with contributions from $\pi^0, \eta,
\omega, \eta'$, and $\phi$ and the mass distribution for virtual
direct photons $f_\mathrm{direct}$ are separately normalized to the
data at $m_{ee} < \unit[30]{MeV}$. Then $r$ is extracted by fitting
$f(m_{ee}) = (1-r ) f_\mathrm{cocktail}(m_{ee}) + r
f_\mathrm{direct}(m_{ee})$ in the range $80 < m_{ee} <
\unit[300]{MeV}$ (see Figure~\ref{Fig:mee}b).

\begin{figure}[t]
\centerline{\includegraphics[width=\textwidth]{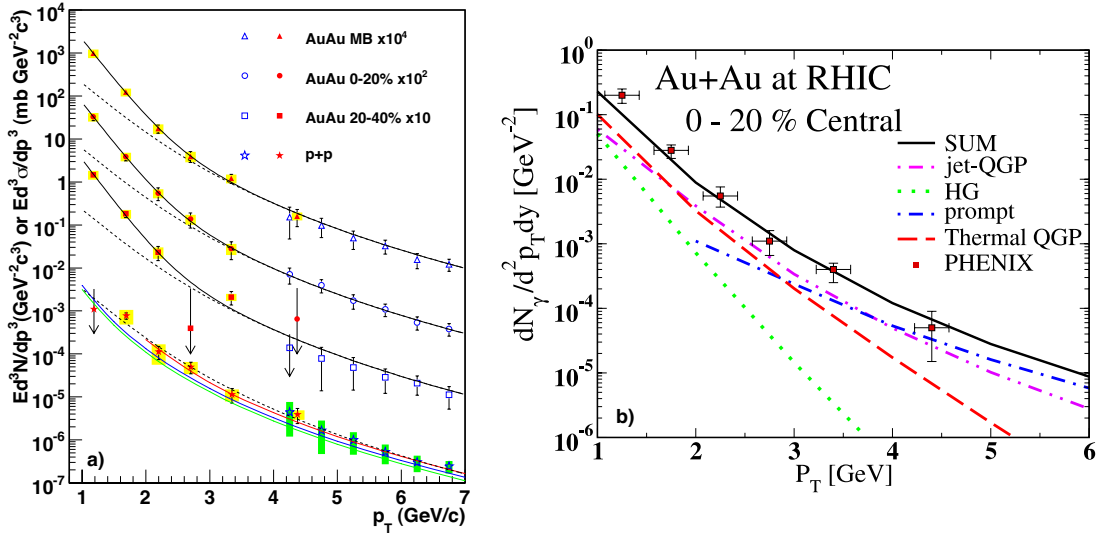}}
\caption{a) Invariant direct-photon cross sections (p+p) and yields
  (Au+Au) at $\sqrt{s_{NN}} = \unit[200]{GeV}$
  \cite{phenix:2008fqa}. The closed symbols are from the internal
  conversion method, the open symbols from EMCal measurements. b)
  Comparison of the direct-photon spectrum in the 20\% most central
  Au+Au collisions with a calculation from
  \cite{Gale:2009gc}}\label{Fig:int_conv_spectra}
\end{figure}

The direct-photon spectra in p+p and Au+Au collisions at
$\sqrt{s_{NN}} = \unit[200]{GeV}$ from the internal conversion method
are shown in Figure~\ref{Fig:int_conv_spectra} along with the EMCal
measurements. The p+p spectrum agrees with the NLO QCD calculation
over the entire range $1 < \pt < \unit[7]{GeV}/c$. The p+p data can be
parameterized with $f_{p+p}(\pt) = A (1 + \pt^2/b)^{-n}$ (dashed
line). For Au+Au the shape of the spectra differs significantly from
the p+p spectrum and yields show a striking enhancement for $\pt
\lesssim \unit[2]{GeV}/c$ with respect to $\langle T_{AB} \rangle
\times f_{p+p}(\pt)$.  A good fit of the Au+Au data can be obtained
with $f_{Au+Au}(\pt) = \langle T_{AB} \rangle \times f_{p+p}(\pt) + B
\exp(-\pt/T)$. The exponential shape of the enhancement in Au+Au is
consistent with the assumption that the excess photons come from a
thermal source. For the 20\% most central Au+Au collisions the
extracted slope parameter is $T = \unit[(221 \pm 23 \pm 18)]{MeV}$.
In hydrodynamical models the initial temperature of the thermalized
quark-gluon plasma is typically 1.5 to 3 times $T$. Thus, if the
excess photons are of thermal origin the measured slope parameter $T$
would indicate an initial temperate well above the critical
temperature for the QGP phase transition.

Figure~\ref{Fig:int_conv_spectra}b shows a comparison of the Au+Au
direct-photon data at low $\pt$ with a calculation which includes all
the direct-photon sources discussed in section \ref{Sec:intro}
\cite{Gale:2009gc}.  The space-time evolution of the fireball is
modeled with ideal hydrodynamics and an equation of state with a
transition from a non-interacting quark-gluon plasma to an chemically
equilibrated hadron resonance gas at a critical temperature of $T_c =
\unit[164]{MeV}$. The contribution from the pre-equilibrium phase is
accounted for by starting the hydro-evolution early at $\tau_0 =
\unit[0.2]{fm/}c$. Assuming full equilibrium at $\tau_0 =
\unit[0.6]{fm/}c$ corresponds to an initial temperature of
$T_\mathrm{initial} = \unit[340]{MeV}$ in this model.  Another piece
of evidence for the creation of a quark-gluon plasma is the fact that
without photons from jet-plasma interactions the data cannot be
described.

\section{Conclusions}
Direct photons at high $\pt$ measured with the PHENIX EMCal played a
crucial role in the discovery of jet quenching at RHIC. Neutral pions
and other hadron in central Au+Au collisions are suppressed whereas
direct photons up to $\pt \approx \unit[14]{GeV}/c$ scale with
$T_\mathrm{AB}$, {\it i.e.}, with the increase of the parton
luminosity per collisions as expected from nuclear geometry. This
strongly supports the interpretation of the high-$\pt$ hadron
suppression as being caused by the energy loss of quark and gluon jets
in the created medium. Direct-photon hadron azimuthal correlations
allow to better constrain the initial jet energy.  The correlation
data were compared to one particular jet quenching model which was
only tuned to describe the single particle $R_\mathrm{AA}$ and
agreement was found. A breakthrough is the measurement of low-$\pt$
direct photons with the internal conversion method in p+p as well as
in Au+Au collisions. The direct-photon spectrum in central Au+Au
collisions spectra exhibits an enhancement above the scaled p+p
spectrum for $\pt \lesssim \unit[2]{GeV}/c$. The exponential shape of
this enhancement and the slope parameter $T > T_c$ are consistent with
the assumption that thermal photons from a QGP phase contribute
significantly to this enhancement.



\begin{footnotesize}


\bibliographystyle{reygers_klaus}
\bibliography{reygers_klaus}

\end{footnotesize}


\end{document}